\documentclass{article}
\usepackage{tipa}
\usepackage{pifont}

\usepackage{amsmath,amssymb,amsfonts,mathrsfs}
\usepackage[latin1]{inputenc}
\usepackage{colortbl}
\usepackage[english]{babel}
\usepackage{graphicx}
\usepackage{bbm}
\usepackage{dsfont}
\usepackage{yfonts}

\begin{document}

\title{\Large Semi-classical States in Homogeneous Loop Quantum
Cosmology}
\author{ Huahai Tan\\
\small Department of Physics, Tsinghua University, Beijing, 100084,
P.R. China\\ \\
Yongge Ma\\
\small Department of Physics, Beijing Normal University, Beijing,
100875, P.R. China}
\date{}
\maketitle

\begin{abstract}
Semi-classical states in homogeneous loop quantum cosmology (LQC)
are constructed by two different ways. In the first approach, we
firstly construct an exponentiated annihilation operator. Then a
kind of semi-classical (coherent) state is obtained by solving the
eigen-equation of that operator. Moreover, we use these coherent
states to analyze the semi-classical limit of the quantum dynamics.
It turns out that the Hamiltonian constraint operator employed
currently in homogeneous LQC has correct classical limit with
respect to the coherent states. In the second approach, the other
kind of semi-classical state is derived from the mathematical
construction of coherent states for compact Lie groups due to Hall.
\end{abstract}


{PACS number(s): 04.60.Pp, 04.60.Kz, 98.80.Qc}

\section{Introduction}
As a candidate of the quantum gravity theory, loop quantum gravity
is noticeable with its background independency
\cite{lecture,AL,HHM,rovelli}. However, the semiclassical analysis
of the theory is still a crucial and open issue \cite{ma}. As
minisuperspace models, loop quantum cosmology (LQC) carries out
quantization by mimicking the construction of loop quantum gravity
\cite{boj}. The symmetry-reduced models provide a mathematically
simple arena to test the ideas and constructions in the full theory.
In the spatially homogeneous and isotropic model, semiclassical
states have been proposed to test the quantum dynamical property of
the theory \cite{ashtekar01}. Moreover, in the light of the quantum
resolution of classical big bang singularity \cite{bojowald},
semiclassical states are currently used to understand the quantum
evolution of the universe across the deep Planck regime \cite{APS}.
However, whether the above results are still robust in more
complicated cases is a crucial question. It is thus desirable to do
similar semiclassical analysis in models with less symmetries, such
as the homogeneous (non-isotropic) LQC formulated by Bojowald
\cite{bojowald01}. To this aim, one still needs suitable
semiclassical states in the homogeneous sector. On the other hand,
the semiclassical analysis of the dynamics in homogeneous LQC has
not been carried out. This is a crucial theoretical criterion for
the correctness of the quantum dynamics.

In this paper, we will construct the desired semi-classical states
in the diagonal homogeneous model with vanishing intrinsic curvature
through two independent ways. Recall that, the coherent states in
the homogeneous and isotropic model arise from the construction in
the polymer-like representation of a single particle mimicking the
construction in loop quantum gravity \cite{ashtekar02}, since the
former is also a quantum mechanics system with one degree of
freedom. It is demonstrated by Ashtekar et al that the coherent
states in the polymer representation are consistent with those in
the traditional Schr\"{o}dinger representation in the low energy
regime. Thus a first attempt is to generalize the construction to
the homogeneous cosmology. We then use the coherent states
constructed by this approach to test the classical limit of the
Hamiltonian constraint operator proposed by Bojowald in homogeneous
LQC. The result is positive. In addition, a mathematical approach is
developed by Hall to construct coherent states on compact Lie groups
\cite{hall01,hall02}. The construction can be directly applied to
the diagonal homogeneous model whose configuration space is a
submanifold of $SU(2)^3$, which is just a compact Lie group.

The rest of this paper is organized as follows. For readers'
convenience, a few basic elements in the diagonal homogeneous LQC
are collected in section 2. The first kind of coherent state for the
homogeneous model is constructed through Ashtekar's approach in
section 3, where a semiclassical analysis of the quantum dynamics is
also carried out. In section 4, the other kind of coherent state is
obtained by Hall's mathematical approach. In section 5, the results
are summarized and some possible applications of our semiclassical
states are discussed.

\section{Diagonal Homogeneous Model}

We work with the diagonal homogeneous model of LQC with vanishing
spatial curvature~\cite{bojowald01,bojowald02}, i.e. the homogeneous
spin connection $\Gamma^i_a=0$, and an Abelian symmetry group. At
the classical level, the invariant connections can be reduced so
that they depend only on a diagonal matrix,
\begin{equation}
    A_a^i=c_{(I)}\Lambda_I^i\omega_a^I,
\end{equation}
where $\Lambda\in SO(3)$ is a rotation matrix, and $\omega_a^I$
($I=1,2,3$) are left-invariant 1-forms under homogeneous symmetry
group, which act as the background structure. Similarly, the
densitized triads can be reduced as
\begin{equation}
    E_i^a=p^{(I)}\Lambda_i^IX_I^a,
\end{equation}
where $X_I^a$ are the left-invariant densitized vector fields dual
to $\omega_a^I$. So the classical theory of homogeneous cosmology is
reduced to a system with finite (three) degrees of freedom. A
configuration is represented by $(c_1, c_2, c_3)$. The Gaussian and
diffeomorphism constraints are naturally resolved by the symmetric
reduction\footnote[2]{Actually, there exists a residual gauge of the
signatures of $p^I$, caused by fixing the background.
$\mathrm{sgn}p^I$ and $|p^I|$ are gauge invariant variables. But we
may firstly consider $p^I$ as the gauge invariant variables, then
eliminate the residual gauge at last.}. The Poisson bracket between
the fundamental configuration and momentum variables in the phase
space reads $\{c_I, p^J\}=\gamma\kappa\delta_I^J$, where $\gamma$ is
the Barbero-Immirzi parameter and $\kappa$ is the gravitational
constant. After the symmetry reduction, the simplified Hamiltonian
reads
\begin{eqnarray}
  H=-2\kappa^{-1}\gamma^{-2}V\left(\frac{c_1c_2}{p^3}+\frac{c_2c_3}{p^1}+\frac{c_3c_1}{p^2}\right).
  \label{classical_hamiltonian}
\end{eqnarray}
There are no connection operators $\hat{c}^I$ in the quantum theory.
Mimicking the full theory, quantum configuration variables are
holonomies. So, at the quantum level, the configuration space
$\mathcal{C}^3$ consists of
\begin{eqnarray}
  \left\{\left(\exp{(c_1\Lambda_1^i\tau_i)},\exp{(c_2\Lambda_2^i\tau_i)},\exp{(c_3\Lambda_3^i\tau_i)}\right):
  c_I\in\mathds{R}, \Lambda\in SO(3)\right\}.
\end{eqnarray}
It is easy to verify that the configuration space is a submanifold
of $SU(2)^3$, but not a subgroup. Note that the quantum
configuration space $\mathcal{C}^3$ is regarded as a product of the
three copies of the submanifold $\mathcal{C}$ of $SU(2)$. The
gauge-invariant measure on the quantum configuration space is the
product Haar measure of three copies of~$SU(2)$ as
\begin{equation}
    d\mu_H^3(c_1,c_2,c_3)=(2\pi)^{-3}\sin^2(\frac{c_1}{2})\sin^2(\frac{c_2}{2})\sin^2(\frac{c_3}{2})dc_1dc_2dc_3.
    \label{measure}
\end{equation}
Thus, the Hilbert space is constructed as the space of
square-integrable functions,
$\mathcal{H}_S=L^2(\mathcal{C}^3,d\mu_H^3)=\left[L^2(\mathcal{C},d\mu_H)\right]^3=\overline{\mathrm{Cyl}}_S.$
Now consider a subspace $L^2(\mathcal{C},d\mu_H)$. The orthonormal
bases of this space consist of holonomies up to a certain factor,
\begin{eqnarray}
  \langle
  c_I|m_I\rangle=\frac{e^{\frac{i}{2}m_Ic_I}}{\sqrt{2}\sin\frac{c_I}{2}},\label{eigenstate}
\end{eqnarray}
where the factor in the denominator is due to integrability under
the Haar measure, and $m_I$ take values in the collection
$\mathds{N}$ of non-negative numbers. The orthonormal basis in
$\mathcal{H}_S$ reads
\begin{eqnarray}
  |m_1,m_2,m_3\rangle=|m_1\rangle\otimes|m_2\rangle\otimes|m_3\rangle.
\end{eqnarray}
The symmetric states in the symmetric Hilbert space $\mathcal{H}_S$
can be expanded as a finite linear combination of the basis,
\begin{equation}
    |\psi\rangle=\sum_{m_1,m_2,m_3}\psi_{m_1,m_2,m_3}|m_1,m_2,m_3\rangle.
\end{equation}
Since desitized triads are commutable, i.e. $\{p^I,p'^J\}=0$, there
exists a momentum representation. The action of fundamental
operators--- the configuration operator (holonomy) and the desitized
triad operator---on the space of cylindrical functions are defined
by multiplication and derivative respectively. The action of a
configuration operator on the basis of $L^2(\mathcal{C},d\mu_H)$
reads
\begin{eqnarray}
    e^{c_I\Lambda_I}|m_I\rangle=\frac{1}{2}(1-2i\Lambda_I)|m_I+1\rangle+\frac{1}{2}(1+2i\Lambda_I)|m_I-1\rangle,
\end{eqnarray}
and
\begin{eqnarray}
    \cos\frac{c_I}{2}|m_I\rangle&=&\frac{1}{2}(|m_I+1\rangle+|m_I-1\rangle),\\
    \sin\frac{c_I}{2}|m_I\rangle&=&\frac{1}{2i}(|m_I+1\rangle-|m_I-1\rangle).
\end{eqnarray}
Moreover, the densitized triad operator and its action on the basis
read
\begin{eqnarray}
    \hat{p}^I&=&-i\gamma\ell_{\mathrm{Pl}}^2\left(\frac{\partial}{\partial
    c_I}+\frac{1}{2}\textrm{cot}\frac{c_I}{2}\right),\\
    \hat{p}^I|m_I\rangle&=&\frac{1}{2}\gamma\ell_{\mathrm{Pl}}^2m_I|m_I\rangle.\label{momentum_operator}
\end{eqnarray}
The eigen-equation (\ref{momentum_operator}) shows that the bases
(\ref{eigenstate}) are eigenstates of the desitized triad operator.
The volume operator in $\mathcal{H}_S$ is constructed by the
densitized triad operators as
\begin{eqnarray}
     \hat{V}&=&\sqrt{|\hat{p}^1\hat{p}^2\hat{p}^3|},
\end{eqnarray}
and its eigen-equation reads
\begin{eqnarray}
     \hat{V}|m_1,m_2,m_3\rangle&=&\left(\frac{1}{2}\gamma\ell_{\mathrm{Pl}}^2\right)^{3/2}\sqrt{|m_1m_2m_3|}|m_1,m_2,m_3\rangle.
\end{eqnarray}
The eigenvalues of volume operator are measured by the cubic of
Planck length. The Hamiltonian constraint operator (with vanishing
spin connection) is
\begin{eqnarray}
    \hat{H}&=&\!\!\!\!\!32i\gamma^{-3}\kappa^{-1}\ell_{\mathrm{Pl}}^{-2}
    \!\left[\sin\frac{c_1}{2}\cos\frac{c_1}{2}\sin\frac{c_2}{2}\cos\frac{c_2}{2}
    \!\!\left(\sin\frac{c_3}{2}\hat{V}\!\cos\frac{c_3}{2}\!-\!\cos\frac{c_3}{2}\hat{V}\!\sin\frac{c_3}{2}\!\right)\right.\nonumber\\
    &&\!\!\!\!\!+\sin\frac{c_3}{2}\cos\frac{c_3}{2}\sin\frac{c_1}{2}\cos\frac{c_1}{2}
    \left(\sin\frac{c_2}{2}\hat{V}\cos\frac{c_2}{2}-\cos\frac{c_2}{2}\hat{V}\sin\frac{c_2}{2}\right)\nonumber\\
    &&\!\!\!\!\!\left.+\sin\frac{c_2}{2}\cos\frac{c_2}{2}\sin\frac{c_3}{2}\cos\frac{c_3}{2}
    \left(\sin\frac{c_1}{2}\hat{V}\cos\frac{c_1}{2}-\cos\frac{c_1}{2}\hat{V}\sin\frac{c_1}{2}\right)\right].\label{hamilton}
\end{eqnarray}
Its action on the basis reads
\begin{eqnarray}
    \hat{H}|n_1,n_2,n_3\rangle\!=\!\!\!\!\!\!\!\!\!\!
    &&\gamma^{-3}\kappa^{-1}\ell_{\mathrm{Pl}}^{-2}\left[\left(V_{n_1,n_2,n_3+1}-V_{n_1,n_2,n_3-1}\right)
    \left(|n_1+2,n_2+2,n_3\rangle\right.\right.\nonumber\\
    &&\left.-|n_1-2,n_2+2,n_3\rangle-|n_1+2,n_2-2,n_3\rangle+|n_1-2,n_2-2,n_3\rangle\right)\nonumber\\
    &&+(V_{n_1,n_2+1,n_3}-V_{n_1,n_2-1,n_3})(|n_1+2,n_2,n_3+2\rangle\nonumber\\
    &&-|n_1-2,n_2,n_3+2\rangle-|n_1+2,n_2,n_3-2\rangle+|n_1-2,n_2,n_3-2\rangle)\nonumber\\
    &&+(V_{n_1+1,n_2,n_3}-V_{n_1-1,n_2,n_3})(|n_1,n_2+2,n_3+2\rangle\nonumber\\
    &&-|n_1,n_2-2,n_3+2\rangle-|n_1,n_2+2,n_2-2\rangle\nonumber\\
    &&+|n_1,n_2-2,n_3-3)].
\end{eqnarray}

\section{Coherent States and Semiclassical Analysis}

Now we begin to construct coherent states in the kinematical Hilbert
space of the homogeneous LQC by the annihilation operator approach
following the idea of Ashtekar et al. We then check whether the
semi-classical limit of the Hamiltonian constraint operator with
respect to these states is consistent with the classical Hamiltonian
constraint~(\ref{classical_hamiltonian}).

The \emph{coherent states} for quantum mechanical systems are
defined as those states satisfying following conditions:
\begin{enumerate}
    \item They belong to the Hilbert space, but every coherent state can be labeled with a point in classical phase space.
          In this sense, classical mechanics can be considered as a
          quantum system that only includes all the coherent states.
    \item Ehrenfest theorem, i.e. the expectation values of the elementary (configuration and momentum) operators
    (and of their commutators divided by $i\hbar$,
          respectively) under a coherent state labeled by a classical phase agree with,
          up to leading order in $\hbar$, the values of corresponding elementary functions (and of their Poisson brackets, respectively)
          at that phase.
    \item The uncertainty relation of the elementary observables under coherent states is
          minimum, e.g., $\Delta x\Delta p=\hbar/2$ in one dimensional single particle quantum mechanics.
\end{enumerate}
For a quantum theory with some classical theory as its classical
limit, one would expect that there exist enough semiclassical
(coherent) states which can represent all the classical solutions.
As we have seen, the framework of loop quantum cosmology (or loop
quantum gravity) looks disparate from that of conventional quantum
mechanics (or quantum field theory). Hence it becomes a crucial task
to check its classical limit by constructing semiclassical states.
In general case, the eigenstates of annihilation operator $\hat{a}$
in quantum mechanics satisfy the above conditions and are usually
also used as a definition of coherent states by physicists.
Following the idea of Ashtekar et al in the construction of coherent
states for a single particle system\cite{ashtekar02}, we will
construct coherent states in the light of an exponentiated
annihilation operator in our diagonal homogeneous model.

Recall that the kernel of the Hamiltonian constraint operator does
not belong to the kinematic Hilbert space $\mathcal{H}_S$, because
they are not normalizable states in general. So the physical states
should be in their natural home $\textrm{Cyl}_S^*$, the algebraic
dual of the symmetric cylindrical function space $\mathrm{Cyl}_S$.
One has a natural inclusion, the Gel'fand triplet,
\begin{equation}
    \mathrm{Cyl}_S\subset\mathcal{H}_S\subset\mathrm{Cyl}_S^\ast.\nonumber
\end{equation}
From the viewpoint of the polymer representation of a single
particle, the Schr\"odinger representation depicts the low energy
physics of a quantum mechanics system, while the intersection of the
polymer Hilbert space $\mathcal{H}_S$ and the Schr\"odinger Hilbert
space contains only the zero element. So, to study the classical
limit of our LQC, one should consider $\mathrm{Cyl}_S^\ast$ which
contains both polymer states and Schr\"odinger states. It is
necessary to ask the question whether $\textrm{Cyl}_S^*$ is big
enough to contain desired coherent states. To answer this question,
one should firstly try to construct the creation operator
$\hat{a}^\dagger$ on $\mathrm{Cyl}_S^\ast$ and treat its
eigenvectors as coherent states. However, the flaw of
$\mathrm{Cyl}_S^\ast$ is that there is no natural inner product on
it. Thus we are not able to calculate the expectation values of
quantum operators. We notice that an arbitrary element in
$\mathrm{Cyl}_S^\ast$ can be formally written as
$$(\Psi|=\sum_{m_1,m_2,m_3}\Psi_{m_1,m_2,m_3}(m_1,m_2,m_3|.$$
Thus one may extract a combination of certain finite terms in the
expression of $(\Psi|$ as an element projected from
$\mathrm{Cyl}^\ast_S$ into $\mathrm{Cyl}_S$. The projected states
are called the ``shadows" of $(\Psi|$ in $\mathrm{Cyl}_S$ in Ref.
\cite{ashtekar02}. The action of a coherent state on a cylindrical
function, which is a linear combination of a finite collection of
$|m_1,m_2,m_3\rangle$, in $\mathrm{Cyl}_{S}$ is the same as the
action of its shadow $\langle\Psi^{\mathrm{shad}}|$ on the same
collection of basis. Each shadow state only captures a part of the
information of $(\Psi|$ contained in $\mathrm{Cyl}_S$, but the
collection of all its shadows is able to determine the complete
properties of the full state.

Now we consider the Gel'fand triplet for one copy of the symmetric
Hilbert space,
$\mathrm{Cyl}_{S,I}\subset\mathcal{H}_{S,I}=L^2(\mathcal{C},d\mu_H)\subset\mathrm{Cyl}^\ast_{S,I}.$
Formally, we may consider a dimensionless "annihilation operator"
\begin{equation}
\hat{a}^I=\frac{\sqrt{2}}{\gamma\ell_{Pl}^2d}(\hat{p}^I-i\frac{\gamma\ell_{Pl}^2d^2}{4}\hat{c}_I),\nonumber
\end{equation}
where $p^I$ has the dimension of $[L]^2$, $c_I$ is dimensionless,
and $d$ is a dimensionless quantity usually representing the
tolerance scaled by $\ell_{Pl}^2$. Note that in Schr\"odinger
quantum mechanics $\hat{a}^I$ has the same eigenstates as those of
the conventional annihilation operator in the configuration
representation. Coherent states are supposed to be the eigenvectors
of the creation operator $\hat{a}_I^\dagger$ in
$\mathrm{Cyl}_S^\ast$. However, the problem is that there is no
connection operator $\hat{c}_I$ in homogeneous LQC. The key idea of
Ashtekar et al is to use the exponentiated creation operator to
solve the problem. For all real number $\alpha$, using the
Baker-Hausdorff-Campbell identity, we have
\begin{equation}
    e^{\sqrt{2}\alpha(\hat{a}^I)^\dagger}=e^{\frac{2}{\gamma\ell^2_{Pl}}\frac{\alpha}{d}\hat{p}^I}V(\alpha
    d)e^{-\frac{\alpha^2}{2}},\label{baker}
\end{equation}
where $V(\mu)=\exp^{i\frac{\mu}{2}c_I}$ belongs to the Weyl algebra
on $\mathcal{H}_{S,I}$, thus also on $\mathrm{Cyl}_{S,I}^\ast$.
Since $V(\mu)$ is a well-defined operator in $\mathcal{H}_{S,I}$, it
turns out that Eq.(\ref{baker}) is well defined on
$\mathrm{Cyl}_{S,I}^\ast$. The coherent states in
$\mathrm{Cyl}_{S,I}^\ast$ are obtained by solving the eigen-equation
\begin{equation}
    (\Psi_{a^I_0}|\left[e^{\frac{2}{\gamma\ell^2_{Pl}}\frac{\alpha}{d}\hat{p}^I}V(\alpha
    d)e^{-\frac{\alpha^2}{2}}\right]=e^{\sqrt{2}\alpha\bar{a}_0^I}(\Psi_{a_0^I}|.
\end{equation}
The distribution state $(\Psi_{a_0^I}|\in\mathrm{Cyl}_{S,I}^\ast$
can be formally expanded as an infinite summation
$(\Psi_{a_0^I}|=\sum_{m_I}\bar{\Psi}_{a_0^I}(m_I)(m_I|$. Its shadow
in $\mathrm{Cyl}_{S,I}$ is
$|\Psi_{a_0^I}^{\mathrm{shad}}\rangle=\sum_{m_I^j}\Psi_{a_0^I}(m_I^j)|m_I^j\rangle$,
where $j$ is finite. The coefficients $\Psi_{a_0^I}(m_I)$ in the
eigen-equation of the exponentiated creation operator are
constrained by
\begin{equation}
    \bar{\Psi}_{a_0^I}(m_I+\alpha
    d)=\exp\left[\sqrt{2}\alpha\bar{a}_0^I-\frac{\alpha
    m_I}{d}-\frac{\alpha^2}{2}\right]\bar{\Psi}_{a_0^I}(m_I),
\end{equation}
for all $\alpha$. It is easy to verify that the shadow state of
$(\Psi_{a_0^I}|$ which peaks at $(P^I,C_I)$ in the classical phase
space is, up to a normalization constant, written as
\begin{eqnarray}
  |\Psi_{I,a_0^I}^{\textrm{shad}}\rangle=\sum_{m_I}
  \left[e^{-\frac{1}{2d^2}(m_I-M_I)^2}
  e^{-i\frac{C_I}{2}(m_I-M_I)}\right]|m_I\rangle,
\end{eqnarray}
where $P^I=\frac{1}{2}\gamma\ell_{\mathrm{Pl}}^2M^I$ and
$\frac{1}{2}\gamma\ell_{\mathrm{Pl}}^2d$ is the ``tolerance" for the
quantum fluctuation of $p^I$. For the semi-classical case, one has
$d\gg 1$. To simulate the classical behavior, we should specify
$M^I\gg d$, which means large volume compared to the Planck volume,
and $C_I\ll 1$, which means small external curvature, thus late
time. Finally, we arrive at a Gaussian-type shadow state of the
desired coherent state $(\Psi_{a_0}|$ in $\mathrm{Cyl}_S^\ast$ by
directly composing the three copies in $\mathrm{Cyl}_S$,
\begin{eqnarray}
  |\Psi_{a_0}^{\textrm{shad}}\rangle=\sum_{m_1,m_2,m_3}\left[e^{-\frac{1}{2d^2}
  \sum_{I=1}^{3}(m_I-M_I)^2}
  e^{-i\frac{C_I}{2}\sum_{I=1}^{3}(m_I-M_I)}\right]|m_1,m_2,m_3\rangle.\label{shadow_needed}
\end{eqnarray}
Note that, a priori, there is no guarantee that any meaningful
calculation in the framework of homogeneous LQC can isolate
semiclassical states corresponding to the standard coherent states
in Schr\"odinger quantum mechanics. So, the existence of the
Gaussian-type coherent states in $\mathrm{Cyl}_S^\ast$ for
homogeneous LQC is a nontrivial result. Since the coefficients in
the expression (\ref{shadow_needed}) coincide with the coherent
states in Schr\"odinger quantum mechanics, in the semiclassical
regime the relevant physics extracted from the conventional quantum
cosmology could also be obtained from homogeneous LQC. One may take
the viewpoint that in the homogenous models LQC is more fundamental
as it incorporates the underlying discrete nature of quantum
geometry, while the conventional quantum cosmology is a kind of
"coarse-grained" description of the fundamental theory. On the other
hand, it is expected that the singularity problem, which still
exists in conventional quantum cosmology, can also be resolved in
homogeneous LQC. As in the single particle case, our toy model also
provides enlightening implications on full loop quantum gravity how
low energy physics can arise from a suitable semiclassical
treatment.

Now we come to the semiclassical analysis of the homogeneous LQC.
This is a crucial step to check the correctness of the quantum
setting. So our next task is to apply the shadow states
(\ref{shadow_needed}) to compute the `expectation value' of
Hamiltonian constraint operator~(\ref{hamilton}), which is
constructed to understand the semi-classical information as
\begin{equation}
    \frac{(\Psi_{a_0}|\hat{H}|\Psi_{a_0}^{\mathrm{shad}}\rangle}{\langle\Psi_{a_0}^{\mathrm{shad}}|\Psi_{a_0}^{\mathrm{shad}}\rangle}
\end{equation}
for any shadow state. Operated by the Hamiltonian constraint
operator, the three copies in~(\ref{shadow_needed}) will intercross
each other. So the computation process is not so simple as in the
homogeneous isotropic case~\cite{ashtekar01}, whose configuration is
$\mathds{R}^1$. We begin with the norm of the shadow state:
\begin{eqnarray}
  \langle\Psi_{a_0}^{\textrm{shad}}|\Psi_{a_0}^{\textrm{shad}}\rangle=
  (\sqrt{\pi}d)^3\left[1+\mathcal{O}\left(e^{-\pi^2d^2}\right)
  +\mathcal{O}\left(\left(M_I/d\right)^{-2}\right)\right].\label{norm}
\end{eqnarray}
Note that, in order to get Eq.(\ref{norm}) we used the Poisson
resummation formula
\begin{eqnarray}
  \sum_{n}e^{-\epsilon^2(n-N)^2}f(n)=\sum_n\int
  e^{-\epsilon^2(y-N)^2}f(y)e^{2i\pi yn}dy,
\end{eqnarray}
and the method of steepest descent (see the appendix of
\cite{ashtekar01}). They lead to a very useful approximation
formula:
\begin{eqnarray}
  \sum_ne^{-\epsilon^2(n-N)^2}f(n)=\sqrt{\pi}\epsilon^{-1}f(N)
  \left(1+\mathcal{O}\left(e^{-\pi^2/\epsilon^2}\right)+\mathcal{O}\left((N\epsilon)^{-2}\right)\right).\label{approximate}
\end{eqnarray}
Set $\epsilon=1/d$, the calculation of
$(\Psi_{a_0}|\hat{H}|\Psi^{\textrm{shad}}_{a_0}\rangle$ can be
divided into three parts. The difference among them is only an index
permutation, since
\begin{eqnarray}
  &&(\Psi_{a_0}|\hat{H}|\Psi^{\textrm{shad}}_{a_0}\rangle=\gamma^{-3}\kappa^{-1}\ell_{\mathrm{Pl}}^{-2}
  \sum_{m_1,m_2,m_3}\sum_{n_1,n_2,n_3}\nonumber\\
  &&\left[e^{-\frac{\epsilon^2}{2}\sum_{I=1}^{3}
  \left((n_I-M_I)^2+(m_I-M_I)^2\right)}e^{i\sum_{I=1}^{3}\frac{C_I}{2}(n_I-m_I)}\right]\nonumber\\
  &&\times[(V_{n_1,n_2,n_3+1}-V_{n_1,n_2,n_3-1})(\delta_{m_1,n_1+2}\delta_{m_2,n_2+2}\nonumber\\
  &&-\delta_{m_1,n_1-2}\delta_{m_2,n_2+2}-\delta_{m_1,n_1+2}\delta_{m_2,n_2-2}+\delta_{m_1,n_1-2}\delta_{m_2,n_2-2})\delta_{m_3,n_3}\nonumber\\
  &&+(V_{n_1,n_2+1,n_3}-V_{n_1,n_2-1,n_3})(\delta_{m_1,n_1+2}\delta_{m_3,n_3+2}\nonumber\\
  &&-\delta_{m_1,n_1-2}\delta_{m_3,n_3+2}-\delta_{m_1,n_1+2}\delta_{m_3,n_3-2}+\delta_{m_1,n_1-2}\delta_{m_3,n_3-2})\delta_{m_2,n_2}\nonumber\\
  &&+(V_{n_1+1,n_2,n_3}-V_{n_1-1,n_2,n_3})(\delta_{m_2,n_2+2}\delta_{m_3,n_3+2}\nonumber\\
  &&-\delta_{m_2,n_2-2}\delta_{m_3,n_3+2}-\delta_{m_2,n_2+2}\delta_{m_3,n_3-2}+\delta_{m_2,n_2-2}\delta_{m_3,n_3-2})\delta_{m_1,n_1}]\nonumber\\
  &\equiv&\gamma^{-3}\kappa^{-1}\ell_{\mathrm{Pl}}^{-2}
  \sum_{m_1,m_2,m_3}\sum_{n_1,n_2,n_3}(A+B+C)\nonumber\\
  &&\times\left[e^{-\frac{\epsilon^2}{2}\sum_{I=1}^{3}
  \left((n_I-M_I)^2+(m_I-M_I)^2\right)}e^{i\sum_{I=1}^{3}\frac{C_I}{2}(n_I-m_I)}\right]\nonumber.\label{semi1}
\end{eqnarray}
We show the calculation of the first part as follows. The other two
are similar. It is easy to see that
\begin{eqnarray}
  \textrm{Part I}&=&\gamma^{-3}\kappa^{-1}\ell_{\mathrm{Pl}}^{-2}
  \sum_{m_1,m_2,m_3}\sum_{n_1,n_2,n_3}A\nonumber\\
  &&\times\left[e^{-\frac{\epsilon^2}{2}\sum_{I=1}^{3}
  \left((n_I-M_I)^2+(m_I-M_I)^2\right)}e^{i\sum_{I=1}^{3}\frac{C_I}{2}(n_I-m_I)}\right]\nonumber\\
  &=&\gamma^{-3}\kappa^{-1}\ell_{\mathrm{Pl}}^{-2}
  \sum_{n_1,n_2,n_3}e^{-\epsilon^2\sum_{I=1}^{3}(n_I-M_I)^2}\nonumber\\
  &&\times[e^{-i(C_1+C_2)}(V_{n_1-1,n_2-1,n_3+1}-V_{n_1-1,n_2-1,n_3-1})\nonumber\\
  &&-e^{i(C_1-C_2)}(V_{n_1+1,n_2-1,n_3+1}-V_{n_1+1,n_2-1,n_3-1})\nonumber\\
  &&-e^{-i(C_1-C_2)}(V_{n_1-1,n_2+1,n_3+1}-V_{n_1-1,n_2+1,n_3-1})\nonumber\\
  &&+e^{i(C_1+C_2)}(V_{n_1+1,n_2+1,n_3+1}-V_{n_1+1,n_2+1,n_3-1})].\nonumber
\end{eqnarray}
Using Eq.(\ref{approximate}), we get
\begin{eqnarray}
  \textrm{Part I}&=&\gamma^{-3}\kappa^{-1}\ell^{-2}_{\mathrm{Pl}}\left(\sqrt{\pi}d\right)^3V_{M_1,M_2,M_3}
  \left(\sqrt{1+\frac{1}{M_3}}-\sqrt{1-\frac{1}{M_3}}\right)\nonumber\\
  &&\times[e^{-i(C_1+C_2)}\sqrt{1-\frac{1}{M_1}}\sqrt{1-\frac{1}{M_2}}
  -e^{i(C_1-C_2)}\sqrt{1+\frac{1}{M_1}}\sqrt{1-\frac{1}{M_2}}\nonumber\\
  &&-e^{-i(C_1-C_2)}\sqrt{1-\frac{1}{M_1}}\sqrt{1+\frac{1}{M_2}}
  +e^{i(C_1+C_2)}\sqrt{1+\frac{1}{M_1}}\sqrt{1+\frac{1}{M_2}}]\nonumber\\
  &&\times\left[1+\mathcal{O}\left(e^{-\pi^2d^2}\right)
  +\mathcal{O}\left(\left(M_I/d\right)^{-2}\right)\right].\nonumber
\end{eqnarray}
Next, taking account of the following approximations:
$$\sqrt{1+\frac{1}{M_I}}=1+\frac{1}{2M_I}+\mathcal{O}\left(M_I^{-2}\right),$$
and
$$e^{i(C_1+C_2)}=1+i(C_1+C_2)-\frac{1}{2}(C_1+C_2)^2+\mathcal{O}\left((C_1+C_2)^3\right),$$
we get
\begin{eqnarray}
  \textrm{Part I}\!\!\!\!\!\!\!\!\!\!&&=-2\left(\sqrt{\pi}d\right)^3\kappa^{-1}\gamma^{-2}V_{M_1,M_2,M_3}\frac{C_1C_2}{P^3}\nonumber\\
  &&\!\!\!\!\!\!\!\!\!\!\!\!\!\!\!\!\!\!\!\!\!\!\times\!\!\left(1+\mathcal{O}\left((M_I/d)^{-2}\right)
  \!+\!\mathcal{O}\left(M_I^{-2}\right)\!+\!\mathcal{O}\left(M_I^{-1}C_I^{-1}\right)\!+\!\mathcal{O}\left(e^{-\pi^2d^2}\right)
  \!+\!\mathcal{O}\left((C_I)^3\right)\right)\!.
\end{eqnarray}
Finally, we arrive at the semi-classical limit of homogeneous LQC by
composing the three copies together,
\begin{eqnarray}
  &&\frac{(\Psi_{a_0}|\hat{H}|\Psi^{\textrm{shad}}_{a_0}\rangle}
  {\langle\Psi^{\textrm{shad}}_{a_0}|\Psi^{\textrm{shad}}_{a_0}\rangle}
  =-2\kappa^{-1}\gamma^{-2}V_{M_1,M_2,M_3}\left(\frac{C_1C_2}{P^3}+\frac{C_2C_3}{P^1}+\frac{C_3C_1}{P^2}\right)\nonumber\\
  &&\!\!\!\!\!\!\times\!\!\left(1+\!\mathcal{O}\left((M_I/d)^{-2}\right)
  \!+\!\mathcal{O}\left(M_I^{-2}\right)\!+\!\mathcal{O}\left(M_I^{-1}C_I^{-1}\right)\!+\!\mathcal{O}\left(e^{-\pi^2d^2}\right)
  \!+\!\mathcal{O}(C_I)\right).\label{final_expectation}
\end{eqnarray}
Thus, under the conditions of large volume and late time of the
universe one has $M_I\gg d$ and $C_I\ll 1$, which further imply
for our diagonal homogeneous model $M_IC_I\thicksim M_I^{3/4}\gg
1$. For simplicity, let us illustrate the validity of the analogue
of the last formula in homogeneous and isotropic cosmology. From
Eq.(4.1.37) in \cite{willis} it is easy to find that the
multiplication of the corresponding momentum and configuration
variables can be estimated as $pc\thicksim a^2\dot{a}\thicksim
a^{3/2}\thicksim p^{3/4}$, where $a$ denotes the scale factor. A
similar estimation is also valid in the diagonal homogeneous
cosmology. Therefore the `expectation value'
(\ref{final_expectation}) agrees with the classical constraint
(\ref{classical_hamiltonian}) up to some small corrections. Thus
we have proved that the Hamiltonian constraint operator $\hat{H}$
in homogeneous LQC has correct classical limit with respect to the
coherent state $(\Psi|$.

\section{On Hall's Coherent States}

It is shown by Hall in \cite{hall01} that, on a compact Lie group
$G$, the coherent transformation is an isomorphism from the space of
square-integrable functions $L^2(G,d\mu_H)$ to the holomorphic
square-integrable function space $\mathcal{H}(G^\mathds{C})\cap
L^2(G^\mathds{C},d\rho)$, that is
\begin{eqnarray}
  \phi_\mathfrak{g}: &L^2(G,d\mu_H)&\longrightarrow\mathcal{H}(G^\mathds{C})\cap
  L^2(G^\mathds{C},d\rho),\\
  &f(g)&\longmapsto\int f(g)\phi_\mathfrak{g}(g)d\mu_H,
\end{eqnarray}
where $G^{\mathds{C}}$ is the complexification of $G$, $d\mu_H$ and
$d\rho$ are, respectively, the Haar measure on $G$ and the heat
kernel measure on $G^\mathds{C}$. The complexification of $G$ can be
carried out by a suitable complexifier $C$ \cite{thiemann01},
\begin{equation}
\mathfrak{g}=\sum_{n=0}^{\infty}\frac{i^n}{n!}\{g,C\}_{(n)},
\end{equation}
which is a suitable function on the cotangent bundle of $G$, where
the Poisson bracket is naturally defined. Then
$\phi_\mathfrak{g}(g)$ is a coherent state in $L^2(G,d\mu_H)$, whose
subscript $\mathfrak{g}$ presents its peak in the classical phase
space. In the appendix of~\cite{hall01}, the concrete form of a
coherent state in $L^2(G, d\mu_H)$ is constructed as
\begin{eqnarray}
  \phi_\mathfrak{g}(g)=\frac{\overline{\rho_t(g^{-1}\mathfrak{g})}}{\sqrt{\rho_t(g)}}\label{coherent},
\end{eqnarray}
where the heat kernel of the compact Lie group $G$ reads
\begin{eqnarray}
  \rho_t(g)=\sum_{\pi\in\hat{G}}\textrm{dim} V_\pi e^{-\lambda_\pi
  t/2}\chi_\pi(g).\label{heatkernel}
\end{eqnarray}
Note that, there is a natural complex analytic continuation
$\rho_t(\mathfrak{g})$ for $\mathfrak{g}\in G^{\mathds{C}}$.
$\hat{G}$ is the set of irreducible representation equivalence
classes, $\textrm{dim}V_\pi$ is the dimension of the representation
space $V$, $\lambda_\pi$ is the factor of the Casimir operator, i.e.
$\pi(\vartriangle)=-\lambda_\pi I$, and $\chi_\pi(g)$ is the
character of $g$ in representation $\pi$. The convergence
of~(\ref{heatkernel}) is also proved in \cite{hall01}.

For homogeneous LQC, although the configuration space
$\mathcal{C}^3$ is not a group, we can still use the compact group
method to construct coherent states, because all the treatment is
carried out in the compact group $SU(2)^3$. Now, the configuration
is $g=e^{c_1\Lambda_1}\otimes e^{c_2\Lambda_2}\otimes
e^{c_3\Lambda_3}$, and we choose a complexifier as
$C=\frac{1}{8}(p^1)^2\otimes (p^2)^2\otimes (p^3)^2$. Taking account
of the direct product structure of $\mathcal{C}^3$, in the following
we present the calculation only for one copy of them, since the
other two are of the same type. Then the complexification is written
as
\begin{eqnarray}
  \mathfrak{g}&=&\sum_{n=0}^\infty\frac{i^n}{n!}\{g,C\}_{(n)}\nonumber\\
  &=&\sum_{n=0}^\infty\frac{i^n}{n!}(\gamma\kappa\Lambda_Ip^I)^ne^{c_I\Lambda_I}\nonumber\\
  &=&e^{(i\gamma\kappa p^I+c_I)\Lambda_I}.\label{complexify}
\end{eqnarray}
Explicitly, it is a kind of complexification of the quantum
configuration space $\mathcal{C}$ and corresponds to a point
$(c_I,p^I)$ in the classical phase space. For the $SU(2)$ group, the
irreducible representations are labeled by their dimensions
$\textrm{dim}V_j=(2j+1)$ with the Casimir $\lambda_j=j(j+1)$. The
character of a representation reads
\begin{eqnarray}
  \chi_j(g)&=&\frac{\textrm{sin}\left[\frac{(2j+1)}{2}c_I\right]}{\textrm{sin}\frac{c_I}{2}}
  =\frac{e^{i(2j+1)c_I/2}-e^{-i(2j+1)c_I/2}}{e^{ic_I/2}-e^{-ic_I/2}}\nonumber\\
  &=&\frac{1}{\sqrt{2}}\left(|2j+1\rangle-|-2j-1\rangle\right).
\end{eqnarray}
Hence the heat kernel in (\ref{heatkernel}) becomes
\begin{eqnarray}
  \rho_t(g)&=&\sum_j(2j+1)e^{-j(j+1)t/2}
  \frac{\textrm{sin}\left[\frac{(2j+1)}{2}c_I\right]}{\textrm{sin}\frac{c_I}{2}} \label{rho}\\
  &=&\sum_{n=1}^\infty
  \frac{1}{\sqrt{2}}ne^{-(n^2-1)t/8}\left(|n\rangle-|-n\rangle\right)\label{rho1}.
\end{eqnarray}
Note that Eq.(\ref{rho1}) is explicitly an eigenstate of the volume
operator with eigenvalue 0, thus $|p|=0$.
Substituting~(\ref{complexify}) and (\ref{rho}) into
(\ref{coherent}), we get the coherent states:
\begin{eqnarray}
    \phi_\mathfrak{g}(g)=\frac{\sum_{n=1}^\infty
    ne^{-(n^2-1)t/8}\frac{e^{-n\gamma\kappa
    P^I/2}e^{-in(C_I-c_I)/2}-e^{n\gamma\kappa
    P^I/2}e^{in(C_I-c_I)/2}}{e^{-\gamma\kappa
    P^I/2}e^{-i(C_I-c_I)/2}-e^{\gamma\kappa
    P^I/2}e^{i(C_I-c_I)/2}}}{\sqrt{\sum_{m=1}^\infty
    me^{-(m^2-1)t/8}\frac{\sin{(mc_I/2)}}{\sin{(c_I/2)}}}}.\label{hall_coherent}
\end{eqnarray}
where $(C_I,P^I)$ is a point in classical phase space, which is
peaked by $\phi_{\mathfrak{g}}(g)$. However, the structure of
coherent states (\ref{hall_coherent}) is too complicated. Further
work is needed for its applications.

\section{Discussion}

In previous sections, two kinds of coherent states for homogeneous
LQC have been obtained. The coherent states constructed by
Ashtekar's approach lie in the dual space $\textrm{Cyl}_S^\ast$,
which does not carry a natural inner product. However, one may
extract the information of a coherent state from its shadows in
$\mathrm{Cyl}_S$. We further show that the Hamiltonian constraint
operator in homogeneous LQC has correct classical limit with respect
to these shadow states under the large volume and late time limit.
Our result confirms that the current construction of the Hamiltonian
operator (\ref{hamilton}) is feasible at least in theoretical sense.
On the other hand, the coherent states constructed by Hall's compact
Lie group method are rather complicated. For a concrete application
of the last kind of coherent state one still needs to develop some
manageable calculation method for it.

The possible applications of the coherent states constructed in this
paper are fascinating. For instance, one may derive an effective
Hamiltonian expression by specifying certain coherent state peaked
at a particular classical solution for homogeneous cosmology. Recall
that in isotropic LQC, a new repulsive force associated with the
non-perturbative quantum geometry comes into play in Planck regime,
which prevents the formation of the big bang singularity \cite{APS}.
It is desirable to check in anisotropic case if this qualitative
picture would still emerge and if the quantum evolution of the
universe across the deep Planck regime could be simulated
numerically. We leave these open issues for further investigation
\cite{mtz}. As our model has less symmetries than the isotropic one,
we may also try to generalize other remarkable achievements in
isotropic LQC, such as the account of inflation
\cite{boj2}\cite{DH}. Besides its own meaning in quantum cosmology,
the semiclassical analysis in homogeneous model will certainly
provide valuable hints for the investigation of full loop quantum
gravity.

\section*{Acknowledgements} The authors would like to thank one of the referees
for helpful comments on the original version of the manuscript. This
work is supported in part by NSFC(10205002).

\end{document}